# LEVELS OF SPACECRAFT AUTONOMY

**Daniel Baker* and Sean Phillips†‡§**

As autonomous systems are being developed in multiple industries it has been recognized that a phased approach is needed both for technical development and user acceptance. Partially in response, the automotive and aircraft industries have published five or six Levels of Autonomy in an attempt to broadly characterize the amount of autonomy technology that is present in any particular vehicle. We have developed a similar six Levels of Spacecraft Autonomy for similar purposes. Here, we propose extending these ideas to spacecraft constellations, clusters, swarms, ground stations for spacecraft, and strategic collaboration among space and ground assets. We intend these levels of autonomy to be used to characterize and describe high-level capabilities for any given satellite for education and communication with lawmakers, government officials, and the general public. All of the proposed levels mimic those of the automotive and aircraft levels in an attempt to maintain consistency and a common understanding with our intended audiences.

## INTRODUCTION

As autonomous systems are being developed in multiple industries, it has been recognized that a phased approach is needed both for technical development and user acceptance. For most manufactures, each new model of automobile introduces new autonomous features and elements (e.g. adaptive cruise control and automatic parking). At some point these individual features combine to form a higher-level capability (e.g. autonomous speed control and autonomous steering control). From early in space flight's history, most spacecraft systems have incorporated automated processes. Some examples include automatic sun pointing if power levels drop below a predefined level; regularly scheduled ground communication contracts; and transition to a safe mode when anomalies occur. In space, the push for more automation is motivated by several factors: the remoteness of the spacecraft to its human controllers, the inability to repair hardware problems, and the communication time delays and time gaps between the spacecraft and the ground to name a few.

In an attempt to broadly characterize the amount of autonomy technology in a particular system, many industries have published five or six levels of autonomy. Figure 1, shows the six SAE Levels of Driving Automation for automotive vehicles and Figure 2, shows five levels of autonomy for aircraft.[12, 1] Similar levels, typically on a 0-5 scale, have been proposed in other industries, including:

- Drone Flight ([5], [8], [13])

*Aerospace GNC Engineer, Defense, BlueHalo, LLC., Albuquerque, NM, USA.

†Technical Advisor, Space Control Branch (RVSW), Air Force Research Laboratory, Albuquerque, NM, USA.

‡This work is sponsored by the Air Force Research Laboratory (RVSW). The views expressed are those of the author and do not necessarily reflect the official policy or position of the Department of the Air Force, the Department of Defense, or the U.S. government.

§Approved for public release; distribution is unlimited. Public Affairs approval #AFRL-2025-0208.

- Industrial and Manufacturing ([3])
- Robotic Surgery ([14])
- Information Technology and Telecommunications Networks ([7], [6])
- Trusted Autonomy ([10], [11]).

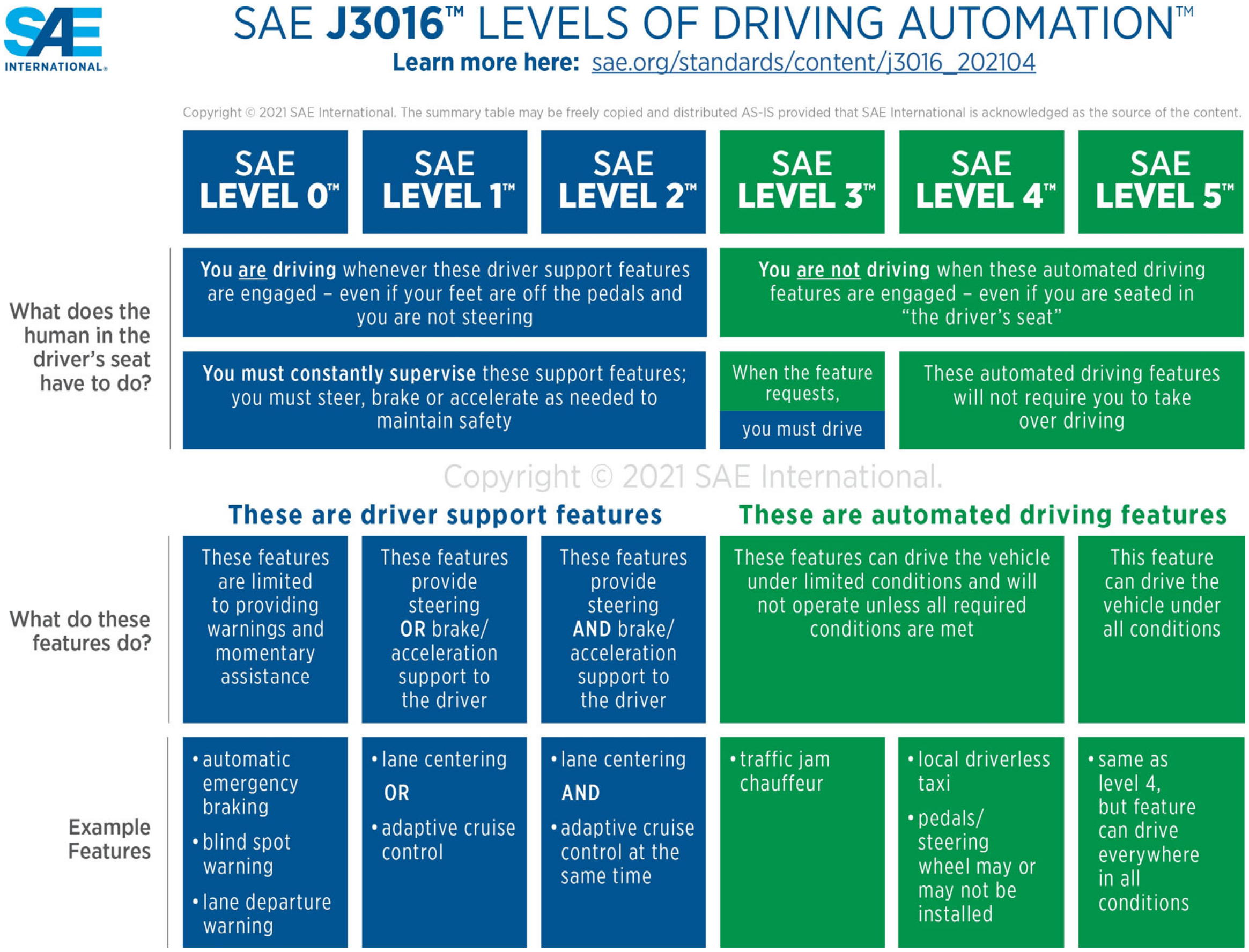

**Figure 1. SAE Levels of Driving Automation**

Following those (and other) examples, in [4] we established an initial set of spacecraft autonomy levels and that work was focused on the autonomy within a single spacecraft. But the situation of autonomy in space is more complicated. In the past, most satellites were single, unique, and purpose built for a specific job. Common busses with the capability of hosting many different payloads are the norm now as are constellations of mass produced satellites. In this work, we describe extensions to our previous work on spacecraft autonomy levels that develop levels for ground segment autonomy, user segment autonomy, and constellation/cluster/swarm autonomy.

Similarly, we intend these additional levels of autonomy to characterize and describe high-level capabilities. Defining autonomy levels in this way maintains consistency with other domains as autonomy capabilities are increasing added to systems. It is a top-down approach to developing autonomy concepts which can be useful in education and communication with lawmakers, government officials, and the general public. We also see this as a launch point for some common understanding and language for future discussions as we work to get a more integrated and comprehensive

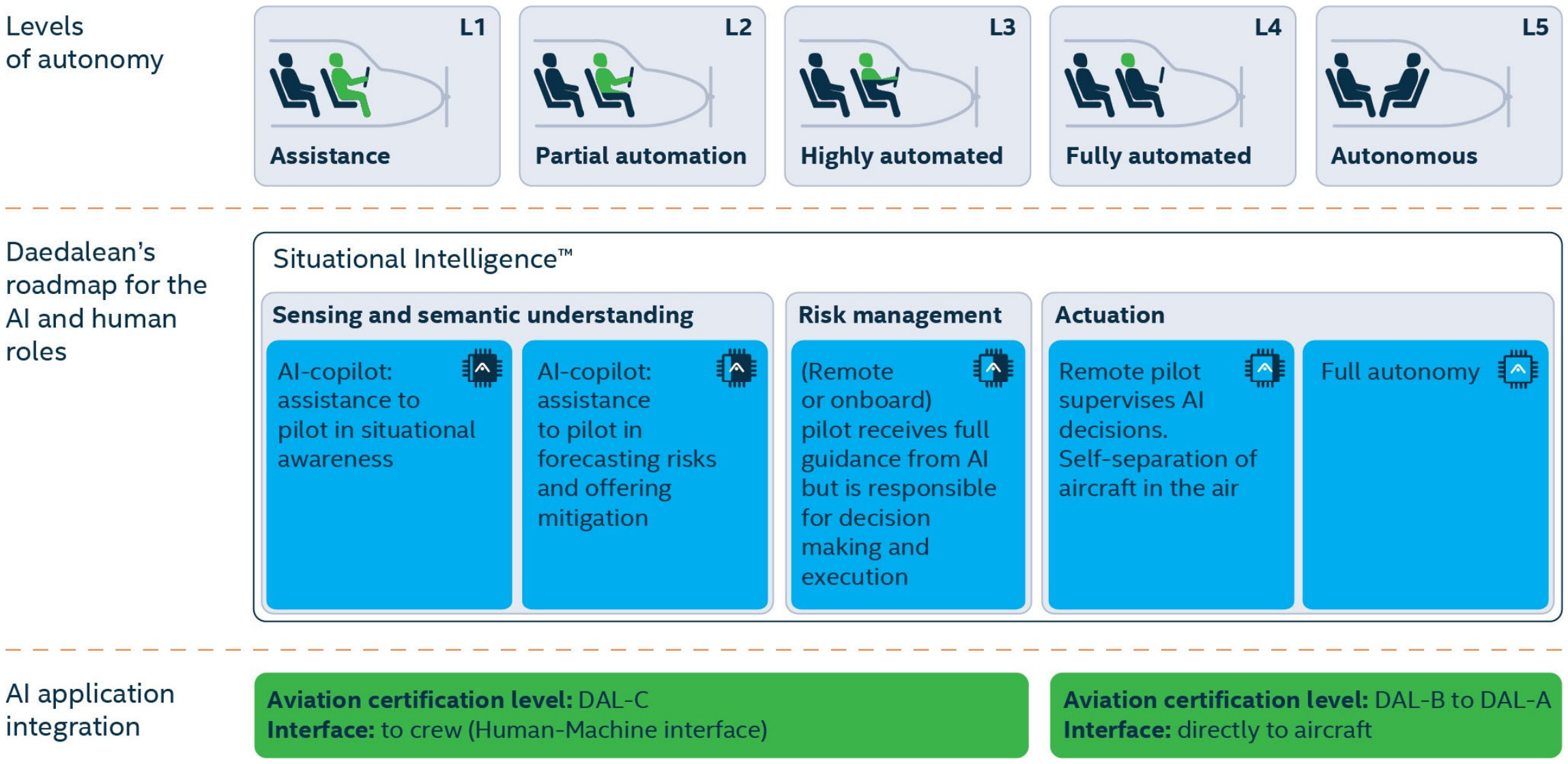

**Figure 2. Aircraft Autonomy Levels**

bottom-up architecture developed.

The remainder of the paper is organized as follows: The next three sections review the previous work, the current levels of spacecraft autonomy, and the needs for expanding the levels. Then we discuss expanding the levels to include autonomy of constellations, clusters, and swarms. The following section then looks at extensions for autonomy within the ground segment including ground station autonomy and after that we discuss tactical vs. strategic autonomy. We end with a summary and our conclusions.

## REVIEW OF PREVIOUS WORK

Even though we have proposed six levels of spacecraft autonomy, we see autonomy as a continuous sliding scale from pure automation of repetitive tasks to complete autonomy. This can also be seen as a scale from dependence (like for a new born baby) to independence (of an adult). For machines it looks like a scale from dependence on to the independence from its system design, given knowledge, and a priori knowledge. Figure 3 summarizes some of the aspects of the two extremes of the scale.

Autonomy means many things to many people. We will not attempt to define autonomy here, but simply recognize that we all need to come to some general understanding of and way of communicating about autonomy. To that end, we are developing a practical framework for autonomy to assist in the design, implementation, use, and test of autonomous systems. We need a usable working description and a way to organize and communicate what, how, and why about the autonomy capability we are discussing.

Autonomy (for humans or machines) invokes the notion of independence. The ability to direct one's own decisions from personal life to interactions with the world. In other words, to control what one does, when one does it, and who one does it with. From the human perspective, it is easy to see this independence play out between children and parents and between workers and managers. For machines, in particular, we primarily think of the system being able to function without (or with little) need for human intervention. Which is back to the basic independence idea. Whether for humans or machines, decision making is a key skill and especially explain-ability of decisions and actions. Rational, understandable, and reasonable decision making is what facilitates trust.

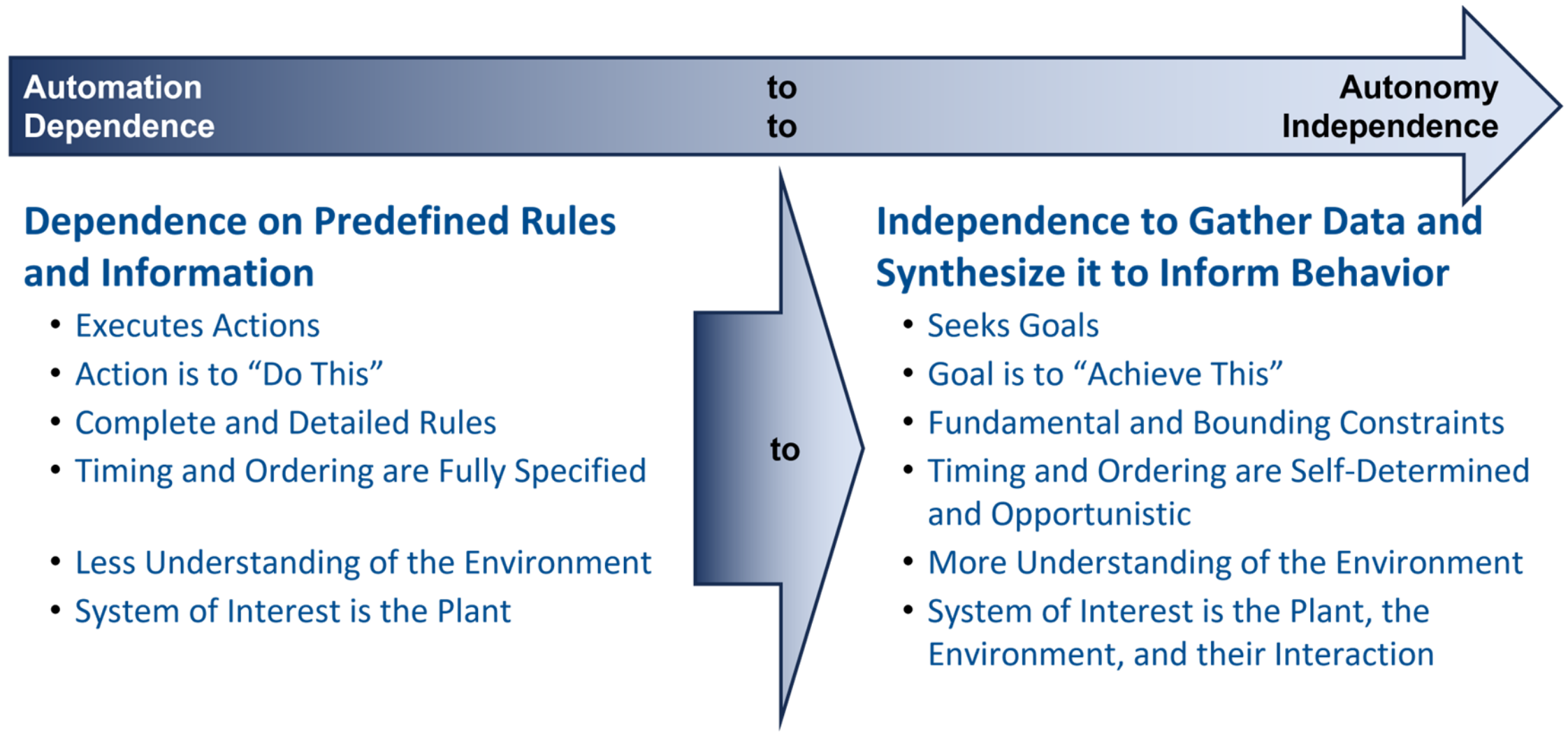

**Figure 3. Extremes of Autonomy on a Sliding Scale**

In [2], Antsaklis and Rahnama make the argument that ”Autonomous systems evolve from control systems ...”. We share this viewpoint and come at it from a slightly different perspective. Autonomy is about getting things done. Machines (systems, things, etc.) need to move to do things (work). The motion of satellites, cars, planes, robots, machines, etc. is governed by physics and described by differential equations. Controls is the engineering discipline for regulating motion and we see autonomy as the engineering discipline for regulating controls. That is, autonomy uses controls to execute actions in order to achieve its (autonomy's) goals. Autonomy is the who, what, when, and where is to be done and controls is how it is done. Autonomy determines what needs doing and controls does the doing.

We have identified four primary (high-level) functions (or purposes) of all autonomy systems: Survival, Success, Collevtive, and Contextualization of Situations (the SSCCs). Next, we describe these four purposes.

**Survival** refers to the fact that all autonomy systems want to act in a way that ensures its own continuation. Humans have this innate survival instinct and autonomy systems should as well. Autonomy cannot achieve its goals if it ceases to operate. Survival includes all of our normal ideas of safety, health, and well being as they apply to both humans and machines. In military situations, survival can also imply undetectability and stealth. In a hierarchical autonomy system, a high-level element could allow/mandate the demise (sacrifice) of a lower-level element for survival of larger system. A simple example of the sacrifice of a lower-level element would be of a fuse blowing to protect the larger circuit. In a military context, an example is the sacrifice of a soldier for the good of the larger campaign, even though the solder has high survival motivations.

**Success** refers to the function of the autonomous system to meet or attempt to meet its mission, goals, and objectives. Its aim to achieve this success is in spite of knowing the system constraints, limitations and potential impacts of decisions. Success gets to the core of the idea of replacing the human by accomplishing the tasks that humans would do. This purpose clearly involves the skills of being able to develop plans, priorities, and tasks within given constraints, boundaries, and authority and being able to execute those devised tasks.

**Collective** is the third function of an autonomy system. No autonomous agent (human, system, or machine) operates in complete isolation from other agents. Autonomy must work with other elements and agents and this implies communication among the agents and elements. The working together can serve several purposes: reporting, commanding, and sharing information and/or tasking. With humans in the loop, this also needs to include human-machine interface design and task load concerns.

**Contextualizing Situations** is maybe the most difficult function of the autonomy system. Here, the autonomous system is trying to replace the human capabilities of taking advantages of new opportunities, anomaly and hazard detection, extrapolation of past experiences to new situations, reasoning and understanding, problem solving, and developing plans for action. These situations can stem from changes in the environment, changes in the machine, ineffective previous plans, and unknown and unforeseen circumstances.

## LEVELS OF SPACECRAFT AUTONOMY

In [4] we developed a familiar six levels (0-5) of spacecraft autonomy as shown in Table 1 and Figure 4. These levels of spacecraft autonomy are intended to characterize and describe high-level capabilities for any given satellite. In following examples of autonomy levels from other sectors (e.g. drone flight [5], [8], [13]; industrial and manufacturing [3]; robotic surgery [14]; and information technology and telecommunications [7], [6]), we distinguish between roles of the autonomy system with those of the human flight controllers (also called operators and mission planners). The human operator's roles include planning, scheduling, coordinating, and monitoring the activities of some desired mission or maneuver for the spacecraft. If the autonomy system is to gradually replace the human then it is to subsume these roles. Also, frequently, a first level breakdown of a spacecraft's systems separates the bus from the payload(s). When we speak of the onboard spacecraft autonomy system, we are referring to the combined capabilities of the bus and payload(s), i.e. the complete spacecraft. A brief description of each level follows.

### Level 0 – Basic Spacecraft Controllability

We begin with Level 0 as a basic functional, controllable and command-able, spacecraft. At Level 0, the only autonomous functions are the spacecraft's systems being able to trigger a transition to safe mode upon the detection of some anomaly or out-of-bounds variable. All spacecraft situational analysis (internal and external) and problem solving is done on the ground. The spacecraft is completely commanded by the ground.

### Level 1 – Ground Systems Assistance

In Level 1, the spacecraft autonomy software has very little ability to act. The autonomy system's function is focused on developing and monitoring its local (internal and external) situational awareness. The spacecraft keeps the ground station updated on its situational awareness. The operators can use other telemetry to confirm that the spacecraft's situational analysis is consistent with their own. The spacecraft is still completely commanded by the ground.

**Table 1. Levels of Spacecraft Autonomy**

| |
|---|
| **Level 0 – Basic Spacecraft Controllability** |
| • Spacecraft has All Basic Systems Operational |
| • Spacecraft Autonomy Limited to Automatic Transitions to Safe Modes when Anomalies Occur |
| • Ground Provides Situational Awareness Analysis and Determines all Spacecraft Actions |
| • Ground Schedules and Initiates All Communications |
| **Level 1 – Ground Systems Assistance** |
| • Spacecraft Onboard Autonomy Keeps Ground Systems Updated on its Local (Internal and External) Situational Awareness |
| • No (or very little) Ability to Act Independently |
| • Other Telemetry used by Ground to Confirm the Spacecraft's Situational Analysis |
| **Level 2 – Advanced Assistance** |
| • Onboard Autonomy Forecasts Risks and Devises Mitigation Actions in Assistance to Ground Systems |
| • Still No Authority to Act on Its Own, Initiation by Ground Systems |
| • Adds Capabilities in Reasoning and Planning (over Level 1) |
| **Level 3 – Partial Automation** |
| • Autonomy now Capable of Acting on its Faults and Risks Mitigation Plans Without Ground Approval |
| • Onboard Autonomy Forecasts Mission Needs and Devises Mission Scenarios in Assistance to Ground Systems |
| • Autonomy Develops Mission-level Course(s) of Action and Presents them to the Ground for Approval to Act |
| **Level 4 – Full Automation** |
| • Autonomy Develops and Executes Mission-level Course(s) of Action and Reports Actions and Results to the Ground |
| • Ground Systems Monitors Autonomy Decisions and Actions on a Regular Schedule |
| • Autonomy Collaborates with Ground Systems in Developing Mission Parameters and Mission Planning |
| **Level 5 – Autonomous** |
| • Autonomy Develops and Executes Mission Parameters, Scenarios, Plans, and Full Course(s) of Action (reporting available on request) |
| • Ground Systems Occasionally Monitors Autonomy Decisions and Actions as Oversight |

## Level 2 – Advanced Assistance

At Level 2 the onboard autonomy system is capable of identifying faults and forecasting risks and devising mitigation plans and actions to address the faults and risks. The autonomy system still has little authority for acting. Its additional capabilities, beyond those of Level 1, lie in reasoning and planning. The mitigation plans can be submitted to the ground system for evaluation, modification, and approval. Any action, beyond Level 1, must be initiated by the operator.

## Level 3 – Partial Automation

A Level 3 autonomy system is now capable of acting on its faults and risks mitigation plans without operator approval. Onboard autonomy can now also forecast mission needs and devises mission scenarios in assistance to ground systems. The autonomy system develops mission-level course(s) of action and presents them to the ground for approval to act. The ground operators monitor the autonomy systems actions and uses the system's mission needs and scenarios as mission planning inputs.

**Level 4 – Full Automation**

At Level 4, spacecraft autonomy develops and executes mission-level course(s) of action and reports its actions and results to the ground. The operators monitor these autonomy decisions and actions on a regular basis. Spacecraft autonomy also now can have the capability to collaborate with ground systems in developing mission parameters and mission planning for new missions and new requirements.

**Level 5 – Autonomous**

Finally at Level 5, the spacecraft develops and executes its own mission parameters, scenarios, plans, and course(s) of action and reports on its activities as requested. Ground systems may occasionally monitor onboard decisions and actions in an oversight role.

An interesting thing to note here is that communication between the spacecraft and the ground changes as the levels change. At the lower levels, the language is very prescriptive from the ground, detailing the commands and telemetry at an almost raw data level. As the levels move higher the vocabulary is more focused on awareness, tasks, and goals ending with a focus on strategies and tactics. Beginning at Level 3, the communication becomes more heavily dictated by the spacecraft.

## EXPANDING THE LEVELS OF AUTONOMY

Space systems can be divided into components, typically called segments. Three common ones are the space segment, ground segment, and the link segment. The space segment includes all space systems operating in space. The ground segment includes ground stations with the capabilities to communicate with the space segment. The link segment operates in the information environment and the electromagnetic operations environment. Sometimes launch and user segments are also delineated. The launch segment includes all the necessary support to get the space segment into space. The user segment includes all of the "users" of the information being supplied by the space segment. Users may be linked directly to the space segment or provided information through ground segment relay. Autonomy can be achieved through levels within each of these segments and the combination leads to a varying degree of autonomy across the spectrum. Here, we will focus on levels of autonomy for the space, ground, and (one example of) user segments.

The levels delineated in our previous work focused on the autonomy system that resides on-board a single satellite (space segment). This leaves several possibilities unexplored. What if the space segment includes more than one satellite in cooperation with each other as with constellations, clusters, or swarms? What about autonomy within ground station systems (ground segment)? What about situations where both ground and space systems can support the user segment? In the following, we will address each of these and refer to constellation, cluster, and swarm autonomy as Case 1, ground segment autonomy as Case 2, and an autonomy within user segment example as Case 3.

In all of these cases, the goal is the same, to replace the human or at least reduce their work load. Due to the limited time and bandwidth for communications between the satellite and the ground, autonomy only within the space segment (Case 1) would be preferable to just autonomy within the ground systems (Case 2). With, Case 3 being the best overall situation. It also interesting to consider that best place to start with adding autonomy to the combined ground-space system might be within the ground station systems. In this situation, algorithms can be tested, adjusted, evaluated and ultimately trusted on a faster cycle than those onboard a satellite. This is a development and test question, not an operational one, and will not be discussed further in this paper.

## LEVELS OF AUTONOMY FOR SPACECRAFT CONSTELLATIONS, CLUSTERS, AND SWARMS

We first continue to address autonomy within the space segment and expand from a single satellite to teams of spacecraft working together. There are differing definitions and distinctions between constellations, clusters, and swarms. Our preference is that a swarm contains 10 or less cooperative spacecraft, with more that 10 constituting a swarm. A constellation is a specific configuration of a swarm where positional flexibility may be more limited than a multi-purpose swarm. For our purposes, in this paper, we will use the terms interchangeably.

Looking first to aviation for examples, in [9], Clough proposes 11 levels of autonomy for uncrewed aerial vehicles (UAVs). He has combined both single UAV and multi-UAV levels within the 11 levels. The first five or six levels (0-4/5) are with respect to a single UAV and levels 5/6-10 are for clusters of UAVs. His dividing line between single and multi-vehicles changes from between levels 4 and 5 in his initial set of levels to between 5 and 6 in his final set of levels. This implies that each individual UAV within a cluster must be at the 4/5 individual level before becoming part of a cluster. This seems overly restrictive, especially for heterogeneous clusters of satellites, where some satellites may not require high levels of autonomy even though they potentially fulfill an important role within the cluster. We prefer to discuss spacecraft swarm autonomy within its own set of levels.

We acknowledge that spacecraft clusters can take various forms. One distinction is around whether the cluster uses centralized versus distributed versus hybrid algorithms. And those three may be applied differently to the tasks of planning versus execution. These distinctions don't change the basic question: how much autonomy exists in the space sector as compared to a human operator in the ground segment. In reviewing the levels of autonomy from Table 2, we see no need to change any of the levels nor their descriptions. As expressed, the levels of autonomy can apply to the onboard, in-space capabilities of either an individual satellite of collectively as a constellation, cluster, or swarm.

## LEVELS OF AUTONOMY FOR THE GROUND SEGMENT

Switching gears, we now address Case 2 where autonomy resides in the ground segment systems themselves without regard to the space segment. The proposed six levels of autonomy for the ground segment as shown in Table 2.

The ground station typically provides three functions as related to the spacecraft:

- Sending commands to the spacecraft
- Sending data to the spacecraft
- Receiving telemetry from the spacecraft

In the next subsections, each of these three functions will be discussed.

### Sending Commands

At Level 1 of the levels of spacecraft autonomy, these commands will be very detailed and time sequential list - move to the way-point, rotate to this attitude, collect 'x' frames of imagery, point solar panels at sun and charge batteries, point to the ground station antenna, download data, etc.

**Table 2. Proposed Levels of Ground Segment Autonomy**

| |
|---|
| **Level 0 – No Ground Systems Autonomy** |
| • Humans use ground-based automation tools to assist in developing spacecraft mission plans and analyzing spacecraft telemetry, but no autonomy exists |
| **Level 1 – Operator Assistance** |
| • Ground Systems Autonomy Keeps Human Operators Updated on Space Segment Status |
| • No (or very little) Ability to Act Independently |
| **Level 2 – Advanced Assistance** |
| • Ground Autonomy Forecasts Risks and Devises Mitigation Actions in Assistance to Human Operators |
| • Still No Authority to Act on Its Own, Initiation by Humans |
| • Adds Capabilities in Reasoning and Planning (over Level 1) |
| **Level 3 – Partial Automation** |
| • Ground Autonomy now Capable of Acting on its Faults and Risks Mitigation Plans Without Human Approval |
| • Autonomy Forecasts Mission Needs and Devises Mission Scenarios in Assistance to Humans |
| • Autonomy Develops Mission-level Course(s) of Action and Presents them to Humans for Approval to Act |
| **Level 4 – Full Automation** |
| • Autonomy Develops and Executes Mission-level Course(s) of Action and Reports Actions and Results to Humans |
| • Humans Monitors Autonomy Decisions and Actions on a Regular Schedule |
| • Autonomy Collaborates with Humans in Developing Mission Parameters and Mission Planning |
| **Level 5 – Autonomous** |
| • Autonomy Develops and Executes Mission Parameters, Scenarios, Plans, and Full Course(s) of Action (reporting available on request) |
| • Humans Occasionally Monitors Autonomy Decisions and Actions as Oversight |

Here, the ground's responsibility is to develop these lists of commands in order to support the mission at hand. As the level of autonomy of the spacecraft increases, the kinds of commands change so that at spacecraft Level 5, the ground system is providing the details of mission directives and operational parameters (what to do) but none of the how to do it details. The ground may also request information about decisions and actions in its oversight role.

In all of these situations, the levels of autonomy of the ground segment is not based on what the interface to the spacecraft is but is based on whether a human or autonomous ground element is the agent doing the work to analyze spacecraft telemetry and decide on what commands to send. With those commands being appropriate to the level of autonomy of the spacecraft.

### Sending Data

The ground also plays the role of providing data or information to the satellite. For example, it could include providing state information from ground-based optical or radar systems or maybe ground-base situation awareness or intelligence information (for spacecraft at higher levels). Whether or not this data comes from human or machine initiation, the role of the ground is that of a sensor system that the spacecraft has access to and this is part of the normal ground systems responsibilities. The levels of autonomy of the ground segment is again based on whether a human or autonomous ground element is the agent doing the work to analyze the data and decide on what information to send.

### Receiving Telemetry

As already alluded to, the ground has the responsibility to receive data from the spacecraft and decide what, if anything, to do with it. This includes the situation already mentioned where the telemetry must be analyzed to determine a next set of commands to send. It also includes the situation where information from the spacecraft needs to be forwarded on to other ground or space assets for their awareness or action. Again here, the levels of autonomy of the ground segment is again based on whether a human or autonomous element is doing the work to analyze the space-based data and decide on what do with it.

## LEVELS OF STRATEGIC AUTONOMY

Finally to Case 3, where autonomy exists both within the ground and space segments. In this case the user segment may want to leverage the tactical capabilities of the autonomy system within the ground and space segments. To address this situation we propose a set of strategic autonomy levels (Table 3).

The levels of autonomy discussed, so far, for the both the ground and space segments, can be considered tactical levels in the broader picture of the military's global battle-space. Each satellite (or swarm) and each ground station, even if fully autonomous themselves, are operating at a tactical level within the battle-space. In this scenario, the user could be the battle-space commander and their staff. A strategic user level autonomy system could be developed to collaborate with and coordinate each of these tactical level (space and ground segment) assets. Each asset has its own view of the situation, it has its own functional and performance capabilities, and each has its own analysis and reasoning capabilities. From a user's perspective, all of these capabilities could to be brought to the table to collaborate on the overall campaign plans and their execution and a strategic level autonomous system within the user segment can do this.

These levels of strategic autonomy could also be called user, mission, or battle-space levels. The levels of strategic autonomy parallel the levels of autonomy for ground systems and spacecraft in that as the levels progress, the user (and associated staff) are relieved from detailed planning and coordination tasks. We do not believe humans are ever completely replaced but that they become part of the group of autonomous agents at the table in the collaboration.

In thinking about these strategic levels, we are not asking either the ground or space segment's assets to change their capabilities or domains or responsibility. The autonomy system within each asset should be focused on that asset. This is why an additional autonomous system is envisioned. One that can work with all of the other autonomous systems/assets - ground, sea, air, space, intelligence, human, etc. in order to pull together a global battle-space view and plan for possibilities. Each other individual autonomous system/asset is responsible for supporting the overall effort with its expertise and with its detailed execution plans.

## CONCLUSION

We have shown how the levels of spacecraft autonomy can be extended, with minor modifications, to include (1) constellations/clusters/swarms of spacecraft, (2) ground segment autonomy, and (3) strategic-level user segment collaborations among ground, space, and other segments. We have attempted to maintain some consistency in level labels and discriminators with these previous examples. We expect these levels to be useful in public relations and marketing materials to be able to provide a basic sense of the capabilities of a particular satellite system. We recognize that the level descriptions contain undefined and flexibly interpreted terms but it is a place to start. It also provides little help to designers or builders of autonomous systems. With the lack of some ability to quantify the autonomy of a spacecraft from a bottoms-up approach, we propose this top-down representation. It will provide a point of departure for future discussions about the specific automated elements that might be required to claim a given level of autonomy.

**Table 3. Proposed Levels of User Segment (Strategic) Autonomy**

| |
|---|
| **Level 0 – No Strategic Autonomy** |
| • Humans use ground-based automation tools to assist in developing global battlespace strategies, but no autonomy exists |
| **Level 1 – User Assistance** |
| • Autonomy Keeps User Updated on Global Situational Awareness |
| • Tactical Asset's Autonomy Collaborate with User to Provide Domain Relevant Data |
| • No (or very little) Ability to Act Independently |
| **Level 2 – Advanced Assistance** |
| • Strategic Autonomy Forecasts Needs and Devises Campaign Scenarios in Assistance to the Commander |
| • Tactical Asset's Autonomy Supports User with Forecasted Risks and Mitigation Actions in Their Domains |
| • Still No Authority to Act on Its Own, Initiation by Humans |
| **Level 3 – Partial Automation** |
| • Strategic Autonomy Develops Full Courses of Action and Battle Plans for User Approval given Campaign Parameters |
| • Tactical Asset's Autonomy Supports User with Forecasting Mission Needs and Mission Scenarios in Their Domains |
| **Level 4 – Full Automation** |
| • Strategic Autonomy Collaborates with the User in Developing Campaign Parameters and Courses of Action and Reports Actions and Results to the User |
| • Tactical Asset's Autonomy Supports User with Mission-level Parameters, Planning, and Course(s) of Action |
| **Level 5 – Autonomous** |
| • Strategic Autonomy Develops and Executes Campaign Parameters, Scenarios, Plans, and Full Courses of Action (reporting available on request) |
| • Tactical Asset's Autonomy Supports User by Developing and Executing Detailed Plans and Course(s) of Action in Their Domains |
| • User Monitors Strategic Autonomy Decisions and Actions as Oversight |

It is this future discussion we hope this work will stimulate. What autonomy elements are required for all/most spacecraft and ground stations? Are there autonomy functions that could be optional? Should governments become involved in regulating minimum requirements and certifications? What operations and/or missions are enhanced (increases in effectiveness and/or performance) with autonomy? What operations and/or missions are enabled (otherwise not possible) with autonomy?

Our future work includes a bottoms-up approach to developing autonomous capabilities at all levels within all of a spacecraft's systems and subsystems.

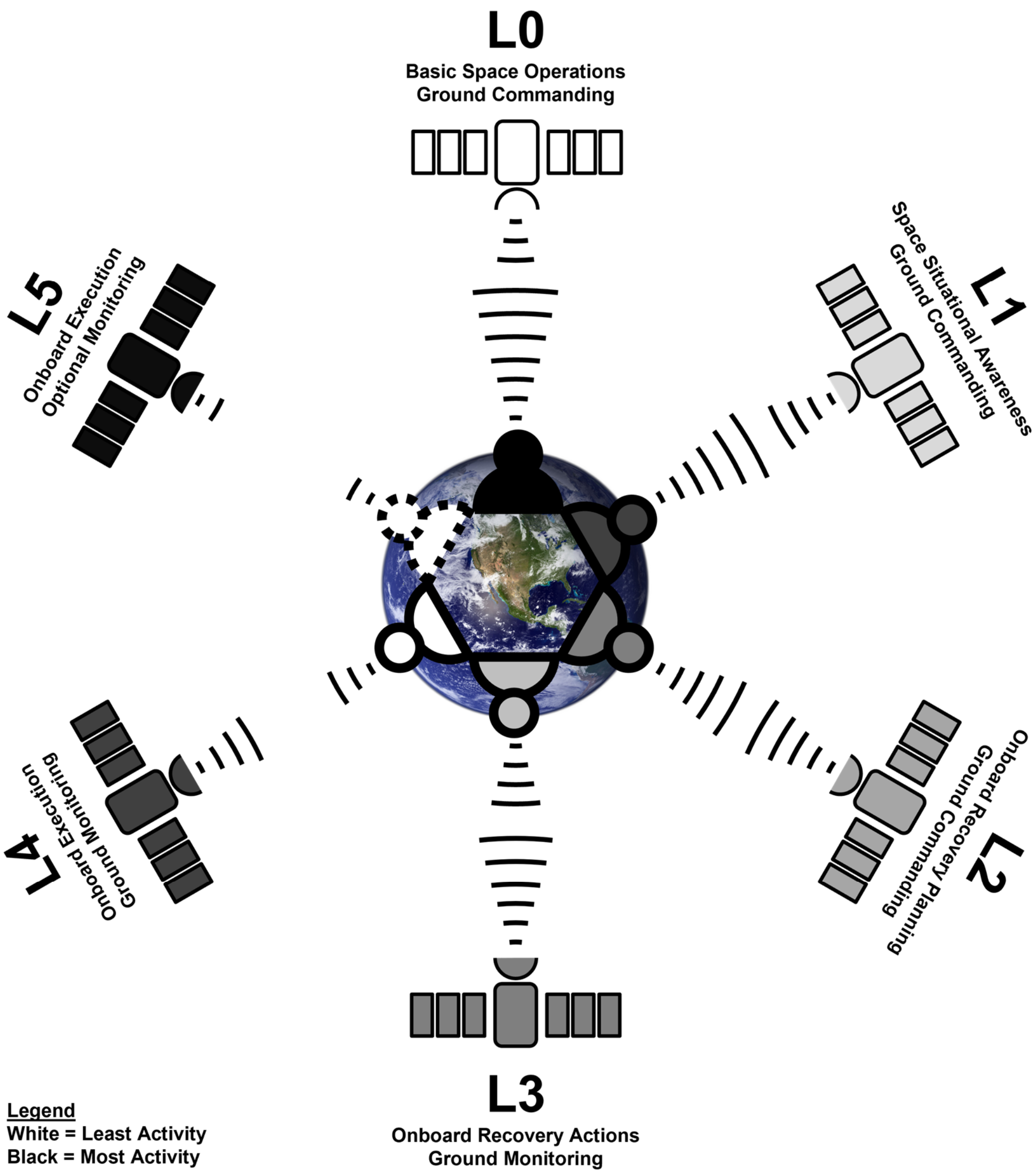

**Figure 4. Levels of Spacecraft Autonomy**